\documentclass[12pt,preprint]{aastex}

\slugcomment{Publications of the Astronomical Society of the Pacific}

\shorttitle{{\em Kepler} Asteroseismology}
\shortauthors{Gilliland, et al.}

\def\rhomean{\langle \rho_* \rangle}
\def\Msun{\,{\rm M}_\odot}
\def\Rsun{\,{\rm R}_\odot}
\begin{document}

\title{{\em Kepler} Asteroseismology Program: \\
Introduction and First Results}

% Place the author information here.  Please hand-code the contact

\author{
Ronald L.~Gilliland,\altaffilmark{1} 
Timothy M.~Brown,\altaffilmark{2} 
J{\o}rgen Christensen-Dalsgaard,\altaffilmark{3} 
Hans Kjeldsen,\altaffilmark{3} 
Conny Aerts,\altaffilmark{4} 
Thierry Appourchaux,\altaffilmark{5} 
Sarbani Basu,\altaffilmark{6} 
Timothy R.~Bedding,\altaffilmark{7} 
William J.~Chaplin,\altaffilmark{8} 
Margarida S.~Cunha,\altaffilmark{9} 
Peter De Cat,\altaffilmark{10} 
Joris De Ridder,\altaffilmark{4} 
Joyce A.~Guzik,\altaffilmark{11} 
Gerald Handler,\altaffilmark{12} 
Steven Kawaler,\altaffilmark{13} 
L\'aszl\'o Kiss,\altaffilmark{7,14} 
Katrien Kolenberg,\altaffilmark{12} 
Donald W.~Kurtz,\altaffilmark{15} 
Travis S.~Metcalfe,\altaffilmark{16} 
Mario J.P.F.G.~Monteiro,\altaffilmark{9} 
Robert Szab\'o,\altaffilmark{14} 
Torben Arentoft,\altaffilmark{3} 
Luis Balona,\altaffilmark{17} 
Jonas Debosscher,\altaffilmark{4} 
Yvonne P.~Elsworth,\altaffilmark{8} 
Pierre-Olivier Quirion,\altaffilmark{3,18} 
Dennis Stello,\altaffilmark{7} 
Juan Carlos Su\'arez,\altaffilmark{19} 
William J.~Borucki,\altaffilmark{20} 
Jon M.~Jenkins,\altaffilmark{21} 
David Koch,\altaffilmark{20} 
Yoji Kondo,\altaffilmark{22} 
David W.~Latham,\altaffilmark{23} 
Jason F.~Rowe,\altaffilmark{20} and 
Jason H.~Steffen\altaffilmark{24} 
}
\altaffiltext{1}{Space Telescope Science Institute, 3700 San Martin Drive, Baltimore, MD 21218, USA; \mbox{gillil@stsci.edu}}
\altaffiltext{2}{Las Cumbres Observatory Global Telescope, Goleta, CA 93117, USA}
\altaffiltext{3}{Department of Physics and Astronomy, Aarhus University, DK-8000 Aarhus C, Denmark}
\altaffiltext{4}{Instituut voor Sterrenkunde, K.U.Leuven, Celestijnenlaan 200 D, 3001, Leuven, Belgium}
\altaffiltext{5}{Institut d'Astrophysique Spatiale, Universit\'{e} Paris XI, B\^{a}timent 121, 91405 Orsay Cedex, France}
\altaffiltext{6}{Astronomy Department, Yale University, P.O. Box 208101, New Haven CT06520, USA}
\altaffiltext{7}{Sydney Institute for Astronomy, School of Physics, University of Sydney, NSW 2006, Australia}
\altaffiltext{8}{School of Physics and Astronomy, University of Birmingham, Birmingham B15 2TT, England UK}
\altaffiltext{9}{Centro de Astrof\'isica da Universidade do Porto, Rua das Estrelas, 4150-762 Porto, Portugal}
\altaffiltext{10}{Royal Observatory of Belgium, Ringlaan 3, B-1180 Brussels, Belgium}
\altaffiltext{11}{Applied Physics Division, LANL, X-2 MS T086, Los Alamos, NM 87545, USA}
\altaffiltext{12}{Institut f\"ur Astronomie, Universit\"at Wien, T\"urkenschanzstrasse 17, A-1180 Wien, Austria}
\altaffiltext{13}{Department of Physics and Astronomy, Iowa State University, Ames, IA 50011, USA}
\altaffiltext{14}{Konkoly Observatory, H-1525, P.O. Box 67. Budapest, Hungary}
\altaffiltext{15}{Jeremiah Horrocks Institute for Astrophysics, University of Central Lancashire, Preston PR1 2HE, UK}
\altaffiltext{16}{High Altitude Observatory and SCD, NCAR, P.O. Box 3000, Boulder, CO 80307, USA}
\altaffiltext{17}{South African Astronomical Observatory, Observatory, 7935 Cape Town, South Africa}
\altaffiltext{18}{CSA, 6767 Boulevard de l'A\'eroport, Saint-Hubert, QC J3Y 8Y9, Canada}
\altaffiltext{19}{Instituto de Astrof\'{\i}sica de Andaluc\'{\i}a, C.S.I.C., Apdo.\ 3004, 18080 Granada, Spain}
\altaffiltext{20}{NASA Ames Research Center, Moffett Field, CA 94035, USA}
\altaffiltext{21}{SETI Institute/NASA Ames Research Center, Moffett Field, CA 94035, USA}
\altaffiltext{22}{NASA Goddard Space Flight Center, Greenbelt, MD 20771, USA}
\altaffiltext{23}{Harvard-Smithsonian Astrophysical Observatory, Cambridge, MA 02138, USA}
\altaffiltext{24}{Fermilab Center for Particle Astrophysics, Batavia, IL 60510, USA}

\begin{abstract}
Asteroseismology involves probing the interiors of stars and quantifying their
global properties, such as radius and age, through observations
of normal modes of oscillation.
The technical requirements for conducting asteroseismology include 
ultra-high precision measured in photometry in parts per million, 
as well as nearly continuous time series over weeks to years, and
cadences rapid enough to sample oscillations with periods as short
as a few minutes.
We report on results from the first 43 days of observations in 
which the unique capabilities of {\em Kepler} in providing a revolutionary
advance in asteroseismology are already well in evidence.
The {\em Kepler} asteroseismology program holds intrinsic importance
in supporting the core planetary search program through greatly enhanced
knowledge of host star properties, and extends well beyond this
to rich applications in stellar astrophysics.

\end{abstract}


\section{INTRODUCTION}

The {\em Kepler Mission} science goals and initial results in the 
core planet-detection and characterization area, as well as the 
mission design and overall performance, are reviewed 
by \citet{bor10}, and \citet{koc10}.

Asteroseismology is sometimes considered as the
stellar analog of helioseismology~\citep{gou96}, being the study of
very low amplitude sound waves that are excited by near-surface,
turbulent convection, leading to normal-mode oscillations in
a natural acoustical cavity.
The Sun, when observed as a star without benefit of spatial resolution
on its surface, shows $\sim$ 30 independent modes with white 
light amplitude of a few parts per million (ppm) and periods 
of 4 -- 8 minutes.   
In many cases stars with stochastically driven oscillations
may deviate strongly from solar size, as in the case of
red giants detailed below.  For our purposes, however, we will
broaden the definition of asteroseismology to also include the
many types of classical variable stars, e.g. Cepheids, famous
for helping to establish the scale of the universe~\citep{hub31}.

There are two primary motivations for performing asteroseismology
with {\em Kepler}, which is primarily a planet-detection
and characterization mission~\citep{bor10}.  First, knowledge of
planet properties is usually limited to first order by knowledge 
of the host star, e.g. {\em Kepler} easily measures the ratio
of planet to star size through transit depths.  Turning this
into an absolute size for the planet requires knowledge of the 
host star size which, in favorable cases, asteroseismology is 
able to provide better than any other approach -- in many cases
radii can be determined to accuracies near 1\%.  Second, the
instrumental characteristics already required for the exquisitely
demanding prime mission (\citealp{bor10}; \citealp{koc10}) can readily support the needs of
seismology at essentially no additional cost or modification, thus
promising strong science returns by allocating a few percent of the
observations to this exciting area of astrophysics.

Oscillations in stars similar to the Sun, which 
comprise the primary set for the planet detections, have periods
of only a few minutes and require use of the Short Cadence
(SC) mode with 58.8-second effective integrations.  Many of the 
classical variables may be studied well with the more standard
Long Cadence (LC) mode, which has 29.4-minute effective integrations.

{\em Kepler} comes at a propitious time for asteroseismology.  
The forerunner missions MOST \citep{mat07}, and especially
CoRoT (\citealp{bag09}; \citealp{mic08}),
have started our journey taking space-based asteroseismology from decades 
of promise to the transformative reality we realistically expect with {\em Kepler}.
The most salient features of the {\em Kepler Mission} for 
asteroseismology are:
(a) a stable platform from which nearly continuous observations
can be made for months to years,
(b) cadences of 1 and 30 minutes which support the vast majority
of asteroseismology cases,
(c) a large 100 square degree field of view providing many stars
of great intrinsic interest,
(d) a huge dynamic range of over a factor of 10,000 in apparent
stellar brightness over which useful asteroseismology (not always
of the same type of variable) can be conducted, and
(e) exquisite precision that in many cases is well under one ppm
for asteroseismology purposes.
Initial data characteristics for SC~\citep{gil10} and LC~\citep{jen10}
have been shown to support results that nearly reach the 
limit of Poisson statistics.

\section{SUPPORT OF EXOPLANETARY SCIENCE}

The {\em Kepler Input Catalog} (KIC)~\citep{koc10} provides knowledge
(30 -- 50\% errors on stellar radii) of
likely stellar properties for $\sim$ 4.5
million stars in the {\em Kepler} field of view at a level
of accuracy necessary to specify the targets to be observed.
One likely application of asteroseismology 
will follow from quantification of stellar radii to more than
an order of magnitude better than this for a few thousand 
giant stars, and several hundred dwarfs, which can then be
used to test the KIC entries, and quite possibly provide the
foundation for deriving generally applicable improvements to
the calibrations enabling redefined entries for the full catalog.

Some 15\% of the KIC entries were not classified, thus no radius
estimates were available to support selection of stars most
optimal for small-planet transit searches.  In Q0 (May 2009) and Q1 (May -- June 2009) a total
of about 10,000 such unclassified stars brighter than Kepler-mag = 13.8
were observed for either 10 and/or 33 days respectively.  An early
application of asteroseismology was to identify stars in this
unclassified set that are obviously red giants, a well-posed
exercise given the quality of {\em Kepler} data~\citep{koc10}, and thus allow
these to be dropped from further observation in favor of bringing
in smaller, and photometrically quieter stars.

At a more basic level asteroseismology can play an important role in
quantifying knowledge of individual planet host candidates.  In particular, by
showing that the stellar radius is significantly greater than the
catalogued value, we can rule out a planetary candidate without the need to
devote precious ground-based resources to measuring radial velocities.  One
such case is the Kepler Object of Interest\footnote{The project
uses these numbers for the internal tracking of candidates \citep{bor10},
which are not synonymous with claimed detections of planets.} (KOI) identified as
KOI-145, whose light curve shows a transit.  The host star
KIC-9904059 has a tabulated radius of just under 4 $\Rsun$, adjusted to 3.6
$\Rsun$ with addition of ground-based classification
spectroscopy\footnote{Based on Spectroscopy Made Easy -- \citealp{val96} --
analysis by Debra Fischer using a spectrum of KOI-145 obtained by Geoff Marcy
at the W. M. Keck Observatory which gives $\rm T_{eff}$ = 4980 $\pm$ 60,
log(g) = 3.47 $\pm$ 0.1, and [Fe/H] = -0.02 $\pm$ 0.06.}. 
This radius allows a transit light curve
solution with a transiting body below 1.6 R$_J$, hence the KOI designation.

\begin{figure}
\epsscale{.6}
\plotone{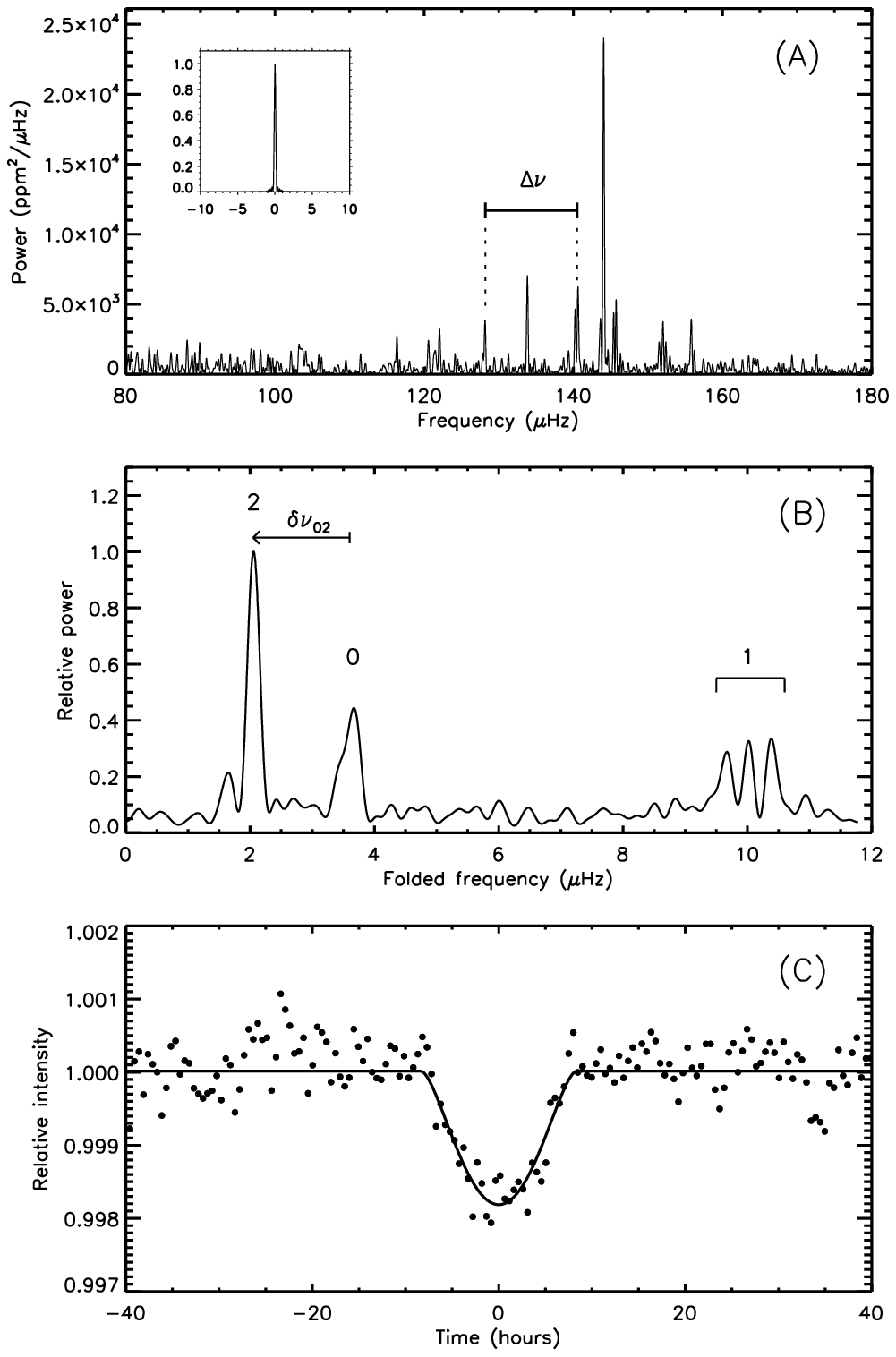}
\caption{
Panel A shows the power spectrum for KOI-145 with the 
signature of solar-like oscillations in a red giant.  This 
shows the power between 80 and 180 $\mu$Hz with an indication
of the large separation parameter, $\Delta \nu$ between two 
consecutive $l$ = 1 modes.  The inset shows the nearly perfect 
spectral window provided by these {\em Kepler} data.
Panel B shows the result of
folding the power spectrum by $\Delta \nu$ which clearly shows the
resulting distribution of $l$ = 2, 0, and 1 (from left) modes corresponding
to the relation given in Eq. 1.
Panel C shows part of the
time series for KOI-145 and a superposed
`transit' light curve solution.  Assuming the stellar radius as implied
by the asteroseismology of R = 5.99 $\pm$ 0.15 $\Rsun$
results in a transiting object size of $\sim$ 0.4 $\Rsun$, well
removed from the planetary realm.
The eclipsing object is likely an M-dwarf star.}
\label{fig:KOI145}
\end{figure}

Figure~\ref{fig:KOI145} shows that KIC-9904059 displays oscillations
characteristic of a red giant star.  The pattern of peaks
in the power spectrum follows the spacing expected from the 
asymptotic relation for low-degree p modes (restoring force
is pressure for sound waves) which may be given 
as (\citealp{tas80}; \citealp{gou86}):
\begin{equation}
\nu_{nl} \approx \Delta \nu (n \, + \, l/2 \, + \, \epsilon) - D_0 l(l+1).
\label{eq:asymptotic}
\end{equation}
Here $\Delta \nu = (2 \int ^R_0 dr/c)^{-1}$ is the so-called large separation, which
corresponds to the
inverse of the sound travel time across the stellar diameter.
In this notation $n$ represents the number of nodes in a radial
direction in the star for the standing sound waves, and $l$ is the
number of nodes around the circumference.  
For stars observed without spatial resolution only $l$ = 0, 1, and 2
modes are typically visible in photometry, since the higher-degree modes that are easily
visible on the Sun average out in disk-integrated measurements.
It can be shown that $\Delta \nu$ scales as the square root of the
stellar mean density, and more precise determinations
of $\rhomean$ follow from using stellar evolution models~\citep{jcd10}.
The $\epsilon$ term captures near-surface effects, while $D_0$ 
represents the so-called small separation, which between $l$ = 0 and
2 we will refer to as:  $\delta_{02}$.  
In main-sequence stars
this small separation is sensitive to the sound speed in the stellar core,
hence the evolving hydrogen content, and helps constrain stellar age.

The asteroseismic solution for KOI-145 yields $\Delta \nu$ = 11.77
$\pm$ 0.07 $\mu$Hz, with maximum power at a frequency
$\nu_{\rm max}$ of 143 $\pm$ 3 $\mu$Hz, and
an amplitude of 50 ppm for the radial modes, all in excellent agreement
with canonical scaling relations.  The $\Delta \nu$ corresponds to
$\rhomean$ = 0.0113 $\pm$ 0.0001 g cm$^{-3}$~\citep{kje08} and, using the spectroscopic constraints
on stellar temperature and metallicity, yields solutions
for mass and radius of 1.72 $\pm$ 0.13 $\Msun$ and 5.99 $\pm$ 0.15 $\Rsun$
following the procedure of \citet{bro09}.

Adopting the larger asteroseismic radius of 5.99 $\Rsun$,
Figure~\ref{fig:KOI145} shows the light curve for KOI-145 (only one eclipse has been seen so far) and an eclipse solution
indicating that the transiting body has a radius of 0.45$^{+0.11}_{-0.07}$ $\Rsun$
assuming an orbital period of 90 days (the minimum currently allowed).
Assuming a longer orbital period allows yet larger values for the
radius of the eclipsing body.  In this case, asteroseismology 
securely sets the radius of the host star and forces the solution 
for the eclipsing object firmly into the stellar domain.
The 3-$\sigma$ lower-limit to the transiting body radius is 3.4 R$_J$.
More detailed analyses await fixing the orbital period
from repeated eclipses.

\begin{figure}
\epsscale{.6}
\plotone{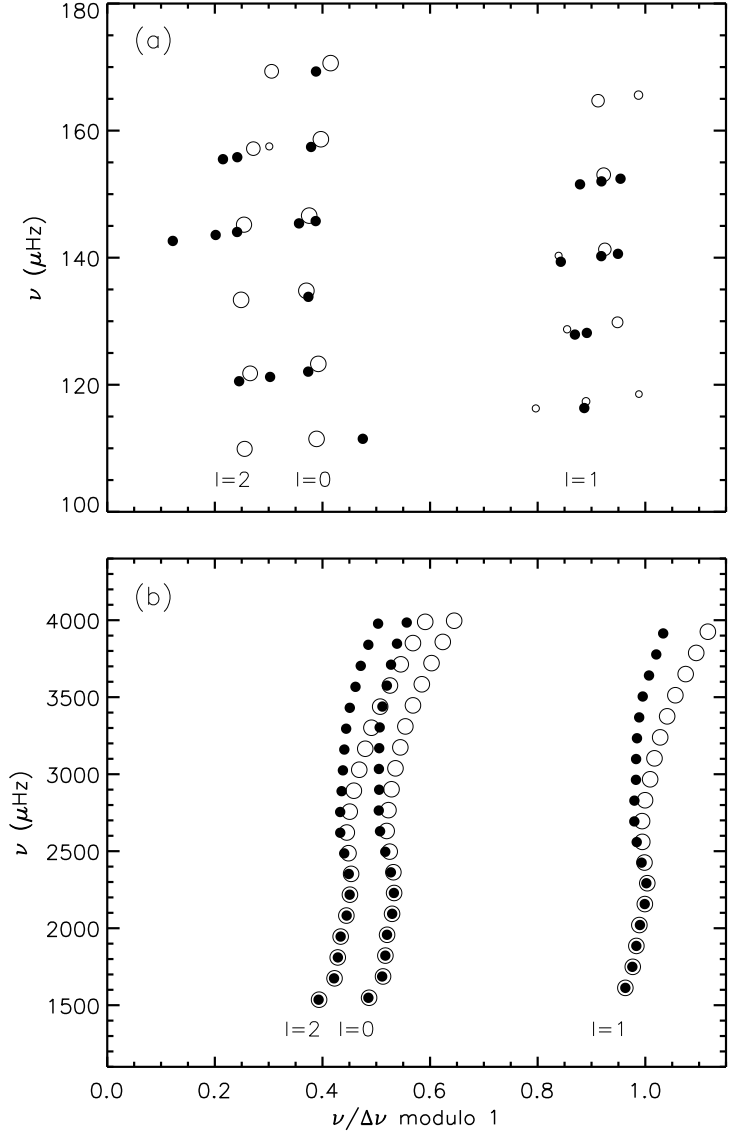}
\caption{
The distribution of low-angular-degree p modes in a so-called
\'echelle diagram, in which individual oscillation frequencies are plotted
against the frequencies modulo the average large frequency spacing, and
with the abscissa scaled to the $\Delta \nu$ units.  Panel a shows results
for KOI-145.  The filled symbols are the observed frequencies and the open
symbols show frequencies from a model in the grid calculated by
\citet{ste09}, with symbol sizes proportional to a simple estimate of the amplitude (see
\citealt{jcd95}).  Panel b shows frequencies for the Sun, with filled
symbols indicating observed frequencies \citep{broo09}, while open symbols
show the frequencies of Model~S \citep{jcd96}.}
\label{fig:rgsurface}
\end{figure}

In Figure~\ref{fig:rgsurface}a the filled symbols show the observed
frequencies in KOI-145 in a so-called \'echelle diagram, in which
frequencies are plotted against the frequencies modulo the average large
frequency spacing (but note that the abscissa is scaled in units of $\Delta
\nu$).  The frequencies were extracted by iterative sine-wave fitting in
the interval $\nu_{max} \, \pm \, 3 \Delta \nu$ down to a level of 19 ppm,
corresponding to 3.3 times the mean noise level in the amplitude spectrum
at high frequency.  The open symbols in Figure~\ref{fig:rgsurface}a show
frequencies from a model in the grid calculated by \citet{ste09}, after
multiplying by a scaling factor of $r=0.9331$ \citep{kje08}.

An interesting feature of Figure~\ref{fig:rgsurface}a is the absence of a
significant offset between observed and model frequencies for the radial
modes.  This contrasts to the Sun, for which there is a long-standing
discrepancy that increases with frequency between observations and standard
solar models, as shown in Figure~\ref{fig:rgsurface}b.  This offset is
known to arise from the inability to adequately model the near-surface
effects \citep{jcd88} and has been an impediment to progress in
asteroseismology for solar-type stars, although an effective empirical
correction has recently been found \citep{kje08} that allows accurate
determination of the mean stellar density from $\Delta \nu$.  As shown in
Figure~\ref{fig:rgsurface}~a, the offset between the measured and
theoretically computed frequencies for $l=0$ modes in KOI-145 is very
small, suggesting some simplification in interpretations for red giants if
this surface term can be generally ignored.  Other factors for red giant
oscillations, however, are already significantly more complex than in the
Sun, as argued in the recent theoretical analysis of \citet{dup09}.  In
particular, less efficient trapping of the modes in the envelope for red
giants on the lower part of the red giant branch can lead to multiple
non-radial modes, especially for $l$ = 1 \citep{dup09}.  We see this in
KOI-145 and also in other red giants observed by {\em Kepler} \citep{bed10}.

\begin{figure}
\epsscale{1.0}
\plotone{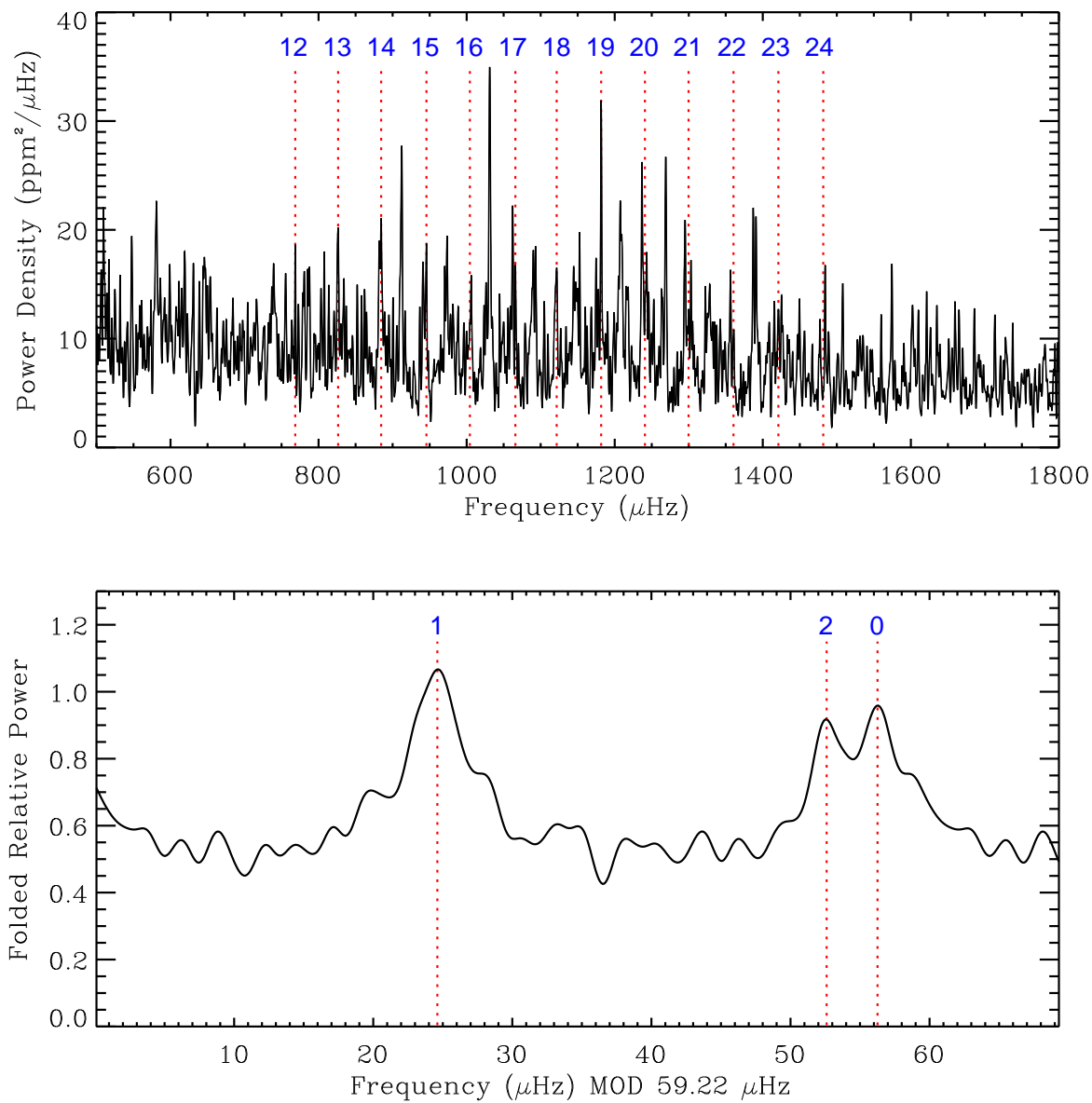}
\caption{
The upper panel shows the power spectrum for the previously
known HAT-P-7 host star based on {\em Kepler} Q0 and Q1 data.  The marked
modes are the radial, $l$ = 0 frequencies labelled with inferred radial order $n$.
The lower panel shows the power spectrum folded by $\Delta \nu$ with the 
$l$ = 1, 2, and 0 ridges (from left) as indicated.}
\label{fig:HATP7}
\end{figure}

The asteroseismic solution for HAT-P-7 by \citet{jcd10} illustrated in
Figure~\ref{fig:HATP7} demonstrates the great potential for refining host star 
properties.  The stellar radius is determined to be
1.99 $\pm$ 0.02 $\Rsun$, an order of magnitude gain in
confidence interval from solutions based on ground-based 
transit light curve solutions by \citet{pal08}, with an age of 1.9 $\pm$ 0.5 Gyr.
Such improvements to host star knowledge, thus ultimately planet
size and density (when radial velocities provide mass) are critically
important for informing studies of planet structure and formation.

\section{INDEPENDENT STELLAR ASTROPHYSICS}

The mission as a whole will benefit from enhanced
knowledge of stellar structure and evolution theory.  This
can best be advanced by challenging theory with detailed 
observations of stellar oscillations across a wide range of stellar
types.  Supporting this goal has led to devoting 0.8\% 
from the available LC target allocations (still a very healthy number
of 1,320 stars) to a broad
array of classical variables.  These LC targets are usually large
stars with characteristic periods of variation of hours, to in 
extreme cases months or even years, and hence are either massive 
and/or evolved stars.  In addition, a set of 1,000 red giants selected
to serve as distant reference stars for astrometry~\citep{mon10} provide 
enticing targets for LC-based asteroseismology.
A number of stellar variables, including, of course, close analogs
of the Sun can only be probed with use of the SC (1-minute) 
observations, on which a cap of 512 targets exists.
Most of these SC slots are now being applied to a survey of
asteroseismology targets rotated through in one-month periods.
After the first year a subset of these surveyed targets will be
selected for more extended observations.  As the mission progresses,
a growing fraction of the SC targets will be used to
follow up planet detections and candidates, for which better
sampling will support transit timing searches for additional planets
in the system~\citep{hol05} and oscillation studies of the host stars.

Asteroseismology with {\em Kepler} is being conducted through
the Kepler Asteroseismic Science 
Consortium (KASC)\footnote{The Kepler Asteroseismic Investigation (KAI)
is managed at a top level by the first 4 authors of this paper.
The next level of authorship comprises the KASC working group chairs,
and members of the KASC Steering Committee. Data for KASC use first
passes through the STScI archive for {\em Kepler}, then if SC is filtered
to remove evidence of any transits, and then is made available to 
the KASC community from the Kepler Asteroseismic Science Operations Centre 
(KASOC) at the Department of Physics and Astronomy, Aarhus University, 
Denmark.  Astronomers wishing to join KASC are welcome to do so 
by following the instructions at:
http://astro.phys.au.dk/KASC/.},
whose $\sim$250 members are organized into working groups
by type of variable star.  
So far data have been available from the first 43 days of the mission
for $\sim$2300 LC targets, some selected by KASC and some being the
astrometric red giants.  The SC data have only been made available for a small number of
targets~\citep{cha10}.
The remainder of this paper reviews the science goals for {\em Kepler}
asteroseismology and
summarizes some of the first results.

\subsection{Solar-like Oscillations}

Stars like the Sun, which have sub-surface convection zones, display a
rich spectrum of oscillations that are predominantly acoustic in
nature\footnote{The solar-like oscillations are driven stochastically and damped
by the vigorous turbulence in the near-surface layers of the
convection zones, meaning the modes are intrinsically stable
see, e.g. \citet{gol77}, and \citet{hou99}.}.
The fact that the numerous excited modes sample different
interior volumes within the stars means that the internal structures
can be probed, and the fundamental stellar parameters constrained, to
levels of detail and precision that would not otherwise be
possible (see, for example, \citealp{gou87}).
Asteroseismic observations of many stars will allow
multiple-point tests to be made of stellar evolution theory and dynamo
theory. They will also allow important constraints to be placed on the
ages and chemical compositions of stars, key information for
constraining the evolution of the galaxy. Furthermore, the
observations permit tests of physics under the exotic conditions found
in stellar interiors, such as those underpinning radiative
opacities, equations of state, and theories of convection.

\subsubsection{Main Sequence and Near Main Sequence Stars}

{\em Kepler} will observe more than 1500 solar-like stars during the initial survey
phase of the asteroseismology program. This will allow the first
extensive ``seismic survey'' to be performed on this region of
the color-magnitude diagram. On completion of the survey, a subset of
50 to 75 solar-like targets will be selected for
longer-term, multi-year observations. These longer datasets will allow
tight constraints to be placed on the internal angular momenta of the
stars, and also enable ``sounding'' of stellar cycles via measurement
of changes to the mode parameters over time \citep{kar09}.

\begin{figure}
\epsscale{0.9}
\plotone{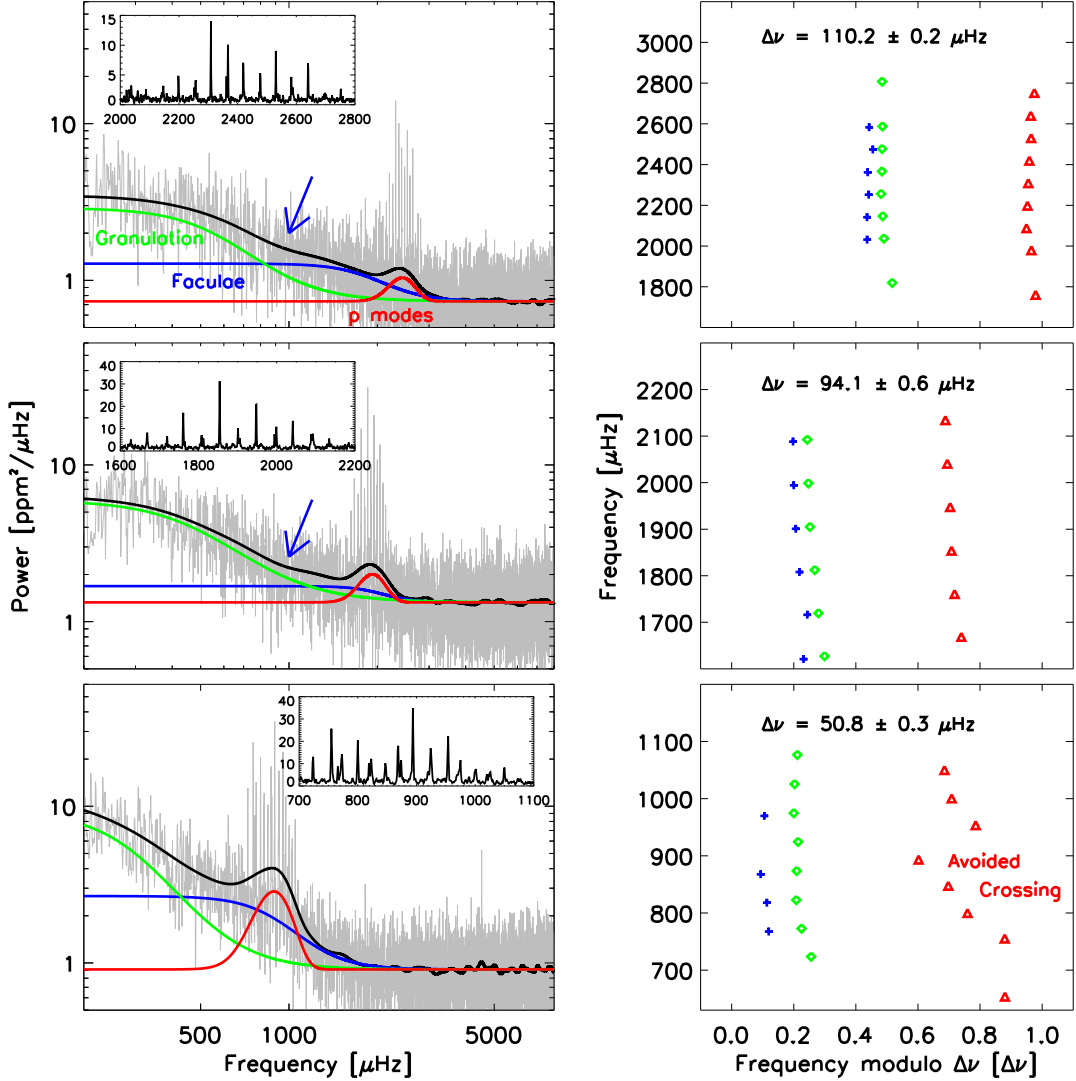}
\caption{
Left-hand panels: Frequency-power spectra of {\em Kepler} photometry of
three solar-like stars (grey) over 200 -- 8000 $\mu$Hz.
The thick black lines show the result
of heavily smoothing the spectra. Fitted estimates of the underlying
power spectral density contribution of p modes, bright faculae
and granulation as labelled in the top left panel are also shown; these
are color coded red, blue and green respectively in the on-line version.
These components sit on
top of a flat contribution from photon shot noise. The arrows mark a
kink in the background power that is caused by the flattening toward
lower frequencies of the facular component. The insets show the
frequency ranges of the most prominent modes.
Right-hand panels: So-called \'echelle plots of individual mode
frequencies. Individual oscillation frequencies have been plotted
against the frequencies modulo the average large frequency spacings
(with the abscissa scaled to units of the large spacing of each
star). The frequencies align in three vertical ridges
that correspond to radial modes ($l=0$, diamonds), dipole modes ($l=1$,
triangles) and quadrupole modes ($l=2$, crosses).}
\label{fig:solarlike}
\end{figure}

Figure~\ref{fig:solarlike} showcases the potential of the {\em Kepler} data for performing
high-quality asteroseismology of solar-like stars from SC data~\citep{cha10}.  The left-hand
panels show frequency-power spectra of three 9th-magnitude, solar temperature targets
observed during Q1. All three stars have a prominent excess of power
showing a rich spectrum of acoustic (p) modes. The
insets show near-regular spacings characteristic of the solar-like
mode spectra, and highlight the excellent S/N observed in the
individual mode peaks. The sharpness of the mode peaks indicates that
the intrinsic damping from the near-surface convection is comparable
to that seen in solar p modes.

The p modes sit on top of a smoothly varying background that rises in
power towards lower frequencies. This background carries signatures of
convection and magnetic activity in the stars.
We see a component that is most likely due to faculae --  bright 
spots on the surface of the stars formed from small-scale, rapidly 
evolving magnetic field. This component is manifest in the spectra of 
the top two stars as a change in the slope of the observed background 
just to the low-frequency side of the p-mode envelope (see arrows). All 
three stars also show higher-amplitude components due to granulation, which
is the characteristic surface pattern of convection.

The near-regularity of the oscillation frequencies allows us to display
them in so-called \'echelle diagrams, in Figure~\ref{fig:solarlike}.  Here, the individual
oscillation frequencies have been plotted against their values modulo
$\Delta \nu$ (the average large frequency spacing -- see
Eq.~\ref{eq:asymptotic}).  The frequencies align in three vertical ridges
that correspond to radial, dipole, and quadrupole modes.
By making use of the individual frequencies and the mean
spacings we are able to constrain the masses and radii of
the stars to within a few percent. The top
two stars are both slightly more massive than the Sun (by about 
5\,\%), and also have larger radii (larger by about 20\,\% and 30\,\% 
respectively).
The bottom star is
again slightly more massive than the Sun (10\,\%), and about twice the radius.
It has evolved off the main sequence, having exhausted the hydrogen in
its core.  The ragged appearance of its dipole-mode ridge
labeled ``Avoided Crossing\footnote{The effects are analogous to avoided crossings of electronic
energy levels in atoms e.g., see \citet{osa75} and \citet{aiz77}.
Here, evolutionary changes to the structure of the deep interior of the star
mean that the characteristic frequencies of modes where buoyancy is
the restoring force have moved into the frequency range occupied by
the high-order acoustic modes. Interactions between acoustic and
buoyancy modes give rise to the avoided crossings, displacing the
frequencies of the dipole modes so that they no longer lie on a smooth
ridge.}" in Figure~\ref{fig:solarlike}, is a
tell-tale indicator of the advanced evolutionary state; the
frequencies are displaced from a near-vertical alignment
because of evolutionary changes to the deep interior structure of the star.

\subsubsection{Red Giants}

Red giants have outer convective regions and are expected to exhibit
stochastic oscillations that are `solar-like' in their general properties
but occur at much lower frequencies (requiring longer time series).
The first firm discovery of solar-like oscillations in a giant was made
using radial velocities by \citet{fra02}. However, it was only recently that the
first unambiguous proof of nonradial oscillations in G and K giants was
obtained, using spaced-based photometry from the CoRoT
satellite~\citep{der09}.  This opened up the field of red giant seismology,
which is particularly interesting because important uncertainties in
internal stellar physics, such as convective overshooting and rotational
mixing, are more pronounced in evolved stars because they accumulate with
age.

\begin{figure}
\epsscale{1.0}
\plotone{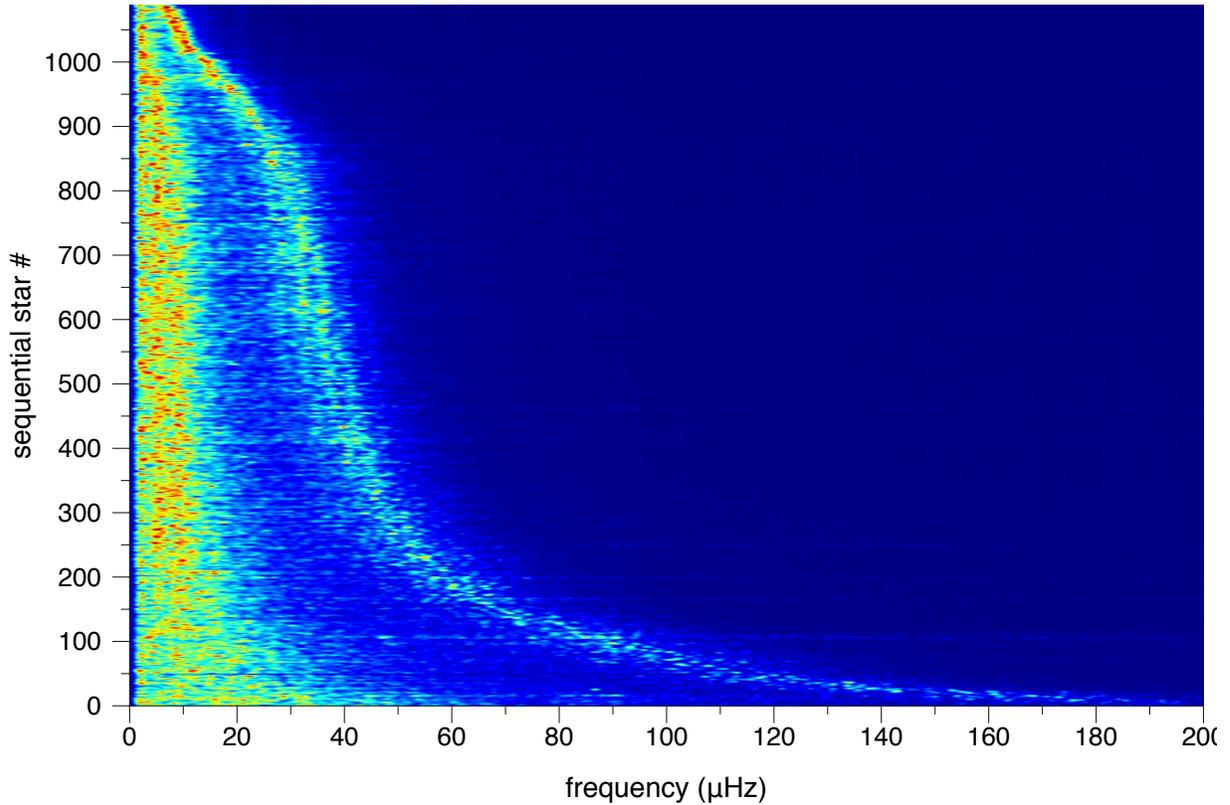}
\caption{
{\em Kepler} red giant power spectra have been stacked
into an image after sorting on $\nu_{\rm max}$ for the oscillation peak.
Larger stars, with lower frequency variations are at the top.
The ever-present slow variations due to stellar granulation are visible
as the band at the left, generally in the 2 - 10 $\mu$Hz range.
The curve from top left to
bottom right traces the oscillations.
Low-mass, low-luminosity giants are clearly visible here with $\nu_{\rm max} > 100$ 
$\mu$Hz~\citep{bed10}.
The slope of the curve is a measure of how fast the evolutionary stage is,
the more horizontal the faster, although selection effects for the sample
also need to be taken into account. The bulk of  
the giants are He-burning stars with a $\nu_{\rm max}$ around 40 $\mu$Hz.
The fact that the top right part of the figure is so uniformly dark illustrates how  
little instrumental noise there is for {\em Kepler}.  By contrast the 
low-Earth orbit for CoRoT imposes extra systematic noise in the general
domain of 160 $\mu$Hz contributing to its lack of results on these smaller,
more rapidly varying and lower-amplitude red giants.}
\label{fig:redgiants}
\end{figure}

The extremely high S/N photometry of the {\em Kepler} observations brings
red giant seismology to the next level.  With the first 43 days of LC data
we were able to detect oscillations with $\nu_{\rm max}$ ranging from 10
$\mu$Hz up to the Nyquist frequency around 280 $\mu$Hz, as shown in Figure~\ref{fig:redgiants}.  The results include the first detection of oscillations in
low-luminosity giants with $\nu_{\rm max} > 100 \, \mu$Hz~\citep{bed10}.
These giants are important for constraining the star-formation rate in the
local disk~\citep{mig09}.  In addition, {\em Kepler} power spectra have such
a low noise level that it is possible to detect $l$=3 modes~\citep{bed10} --
significantly increasing the available asteroseismic information.
The large number of giants that {\em Kepler} continuously monitors for
astrometric purposes during the entire mission will allow pioneering
research on the long-term interaction between oscillations and granulation.
It is also expected that the frequency resolution provided by {\em Kepler}
will ultimately be sufficient to detect rotational splitting in the fastest
rotating giants, and possibly allow the measurement of
frequency variations due to stellar evolution on the red giant branch.

\subsection{Classical Pulsating Stars}

Several working groups within KASC are devoted to the various classes of
classical pulsators.  The goals and first results are discussed below,
ordered roughly from smaller to larger stars (i.e., from shorter to longer
oscillation timescales), giving most emphasis to those areas in which
scientific results have already been possible.

\subsubsection{Compact Pulsators}

The KASC Working Group on compact pulsators will explore the internal
structure and evolution of stars in the late stages of their nuclear evolution
and beyond.  The two main classes of targets are white dwarf stars (the
ultimate fate of solar-mass stars) and hot subdwarf B stars -- less-evolved
stars that are undergoing (or just completing) core helium burning.

While compact stars are among the faintest asteroseismic targets, the science
payoff enabled by continuous short-cadence photometry will be profound.
Pulsations in these stars are far more coherent than for solar-like pulsators,
so extended coverage should provide unprecedented frequency resolution, and
may also reveal very low amplitude modes.

We will address	several important questions about the helium-burning stage
including the relative thickness of the surface hydrogen-rich layer
(i.e. \citealp{char06}), the role and extent of radiative levitation
and diffusion of heavy elements \citep{char08}, and the degree of
internal differential rotation (\citealp{kaw99}; \citealp{kaw05}; and \citealp{char09}).  Even if these
stars rotate extremely slowly, the high frequency resolution of {\em Kepler}
data should reveal rotationally split multiplets.  Asteroseismology can
also probe the role of binary star evolution in producing these hot
subdwarfs \citep{hu08}.

Similar questions about white dwarfs will be addressed, along with probing
properties of the core such as crystallization (e.g. \citealp{met04}; and \citealp{bra05}) and
relic composition changes from nuclear evolution (see \citealp{win08}, and \citealp{fon08}
for recent reviews of white dwarf asteroseismology).

\subsubsection{Rapidly Oscillating Ap Stars}

Rapidly oscillating Ap (or roAp) stars, with masses around twice solar, are
strongly magnetic and chemically peculiar.  They oscillate in high-overtone
p modes, similar to those seen in the Sun (\citealp{kur82}; \citealp{cun07}).  Their
abnormal surface chemical composition results from atomic diffusion, a
physical process common in stars but most evident in Ap stars due
to their extremely stable atmospheres. Since ground-based photometry only reveals
a small number of oscillation modes in roAp stars, long-term continuous
observations from space are needed to detect the large numbers of modes
desirable for asteroseismology.  This has been achieved for a few bright
roAp stars using the MOST satellite, which found oscillation modes down to
the level of 40 $\mu$mag~\citep{hub08}.  Since no roAp stars
are present in the CoRoT fields, {\em Kepler}, 
with its 1 $\mu$mag precision should dramatically increase
the number of detected oscillation modes -- providing strong
constraints on models of diffusion, magnetic fields, and the internal
structure of roAp stars.

\subsubsection{$\delta$ Sct and $\gamma$ Dor Variables}

\begin{figure}
\plotone{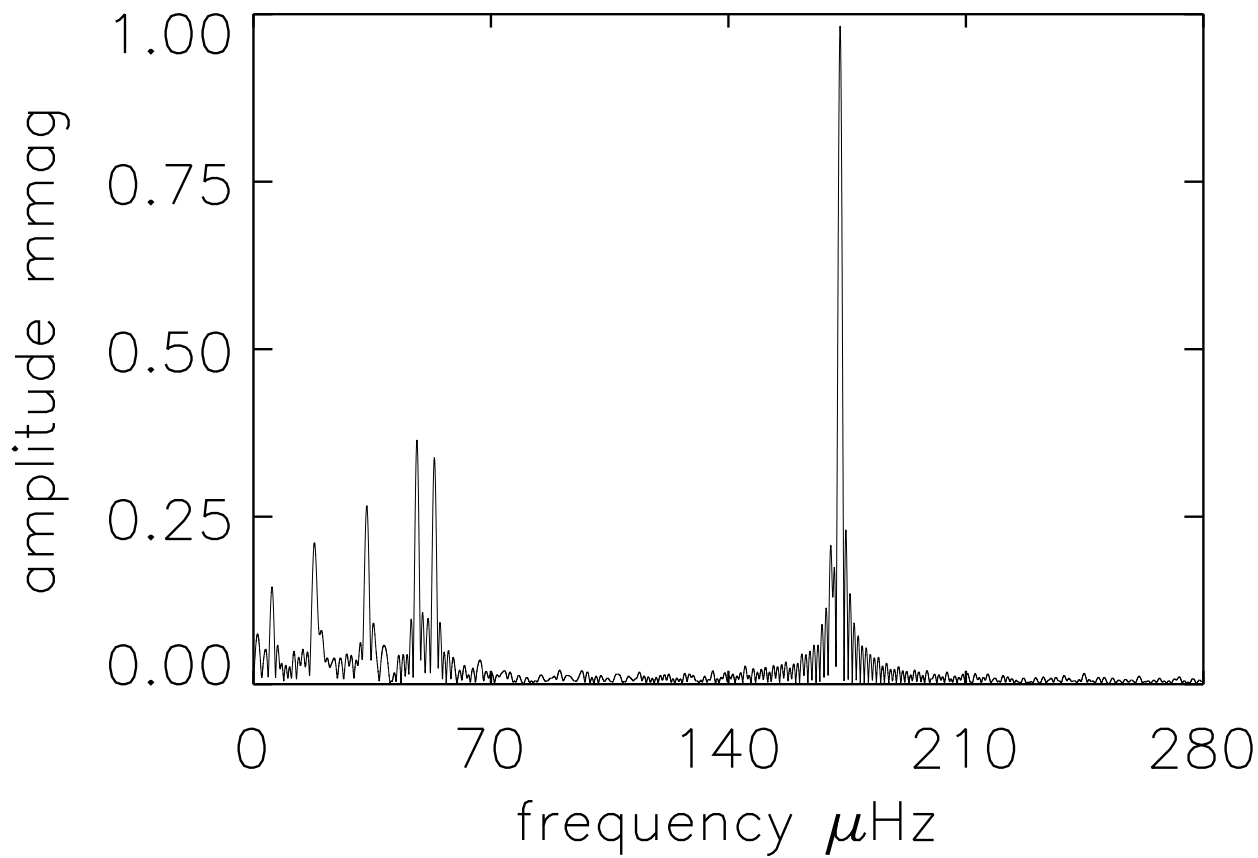}
\caption{
Amplitude spectrum of the $\delta$ Sct -- $\gamma$ Dor star hybrid 
KIC~09775454 where the g-mode pulsations are evident in the $0 - 70$\,$\mu$Hz   
range and the p-mode pulsation is evident at $173$\,$\mu$Hz ($P = 96$\,min).
There are further frequencies in the $\delta$\,Sct range that are not seen at 
this scale. The amplitude spectrum has a high frequency
limit just below the Nyquist frequency of 283\,$\mu$Hz for LC data. In the 
range $70 - 140$\,$\mu$Hz there are no significant peaks. The highest noise
peaks have amplitudes of about 20\,$\mu$mag; the noise level in this
amplitude spectrum is about 5\,$\mu$mag.}
\label{fig:hybrid}
\end{figure}

Stars that exhibit both p modes (pressure-driven) and g modes
(buoyancy-driven) are valuable for asteroseismology because they pulsate
with many simultaneous frequencies.  Theoretical calculations predict a
small overlap in the H-R diagram of the instability region occupied by the
$\gamma$\ Dor stars, which display high-order g modes driven by convective
blocking at the bottom of the envelope convection zone~\citep{guz00}, and
the $\delta$\ Sct stars, in which low-order g and p modes are
excited by the $\kappa$~mechanism in the He\,\textsc{ii} ionization
zone\footnote{The mechanism operates through perturbations to the opacity
$\kappa$ that blocks radiation at the time of compression in
a critical layer in the star, heating the layer and contributing
to driving the oscillation.}.
Among the hundreds of known $\delta$\,Sct and
$\gamma$\,Dor variables, ground- and spaced-based observations have so far
detected only a handful of hybrids~\citep{han09}.  The g modes in
$\gamma$\,Dor stars have low amplitudes and periods of order one day,
making them particularly difficult to detect from the ground.
The {\em Kepler} Q1 data in LC mode have revealed about 40 $\delta$\ Sct
candidates and over 100 $\gamma$\ Dor candidates, among which we find
several hybrid $\delta$ Sct -- $\gamma$ Dor stars~\citep{gri10}.
These are young hydrogen-burning stars with temperatures of
6500--8000\,K and masses of 1.5--2\,M$_{\odot}$.  Figure~\ref{fig:hybrid} shows an
example.

The {\em Kepler} data, in conjunction with spectroscopic 
follow-up observations, will help to unravel theoretical puzzles 
for the hybrid stars.  For example, {\em Kepler} should help us 
identify and explain the frequency of hybrid stars, discover previously 
unknown driving mechanisms, test theoretical predictions of the 
$\delta$\,Sct and $\gamma$\,Dor oscillation frequencies, 
find possible higher-degree modes that fill in frequency gaps, 
and determine whether all hybrids show abundance peculiarities 
similar to Am stars as were noted in previously known hybrids.  
We should also learn about the interior or differential rotation of 
these objects via rotational frequency splittings.  Perhaps just as 
interesting are those stars observed by {\em Kepler} that reside in 
the $\gamma$\,Dor or $\delta$\,Sct instability regions that show no 
significant frequencies.  Indeed, the discovery of photometrically 
constant stars at $\mu$mag precision would be very interesting.  
Studying {\em Kepler's} large sample of stars and looking for trends 
in the data will also help us better 
understand amplitude variation and mode selection.

Another unique possibility arising from the long-baseline, high-precision {\em Kepler} data is 
the ultimate frequency resolution it provides.  The CoRoT mission has recently shown $\delta$ Sct
stars to have hundreds and even thousands of detectable frequencies, 
with degrees up to $l \le 14$~\citep{por09}. However, it has long been known that mode frequencies in 
$\delta$ Sct stars can be so closely spaced that years of data---such as that provided by {\em 
Kepler}---may be needed to resolve them.

\subsubsection{RR Lyr Stars}

RR Lyr stars are evolved low-mass stars that have left the main sequence and
are burning helium in their cores.  Their ``classical" radial oscillations with
large amplitudes make them useful tracers of galactic history and touchstones
for theoretical modelling~\citep{kol10}.  Moreover, like Cepheids, they obey a
period-luminosity-color relation which allows them to play a crucial role as distance
indicators. Most RR Lyr stars pulsate in the radial fundamental mode, the
radial first overtone, or in both modes simultaneously.

\begin{figure}
\plotone{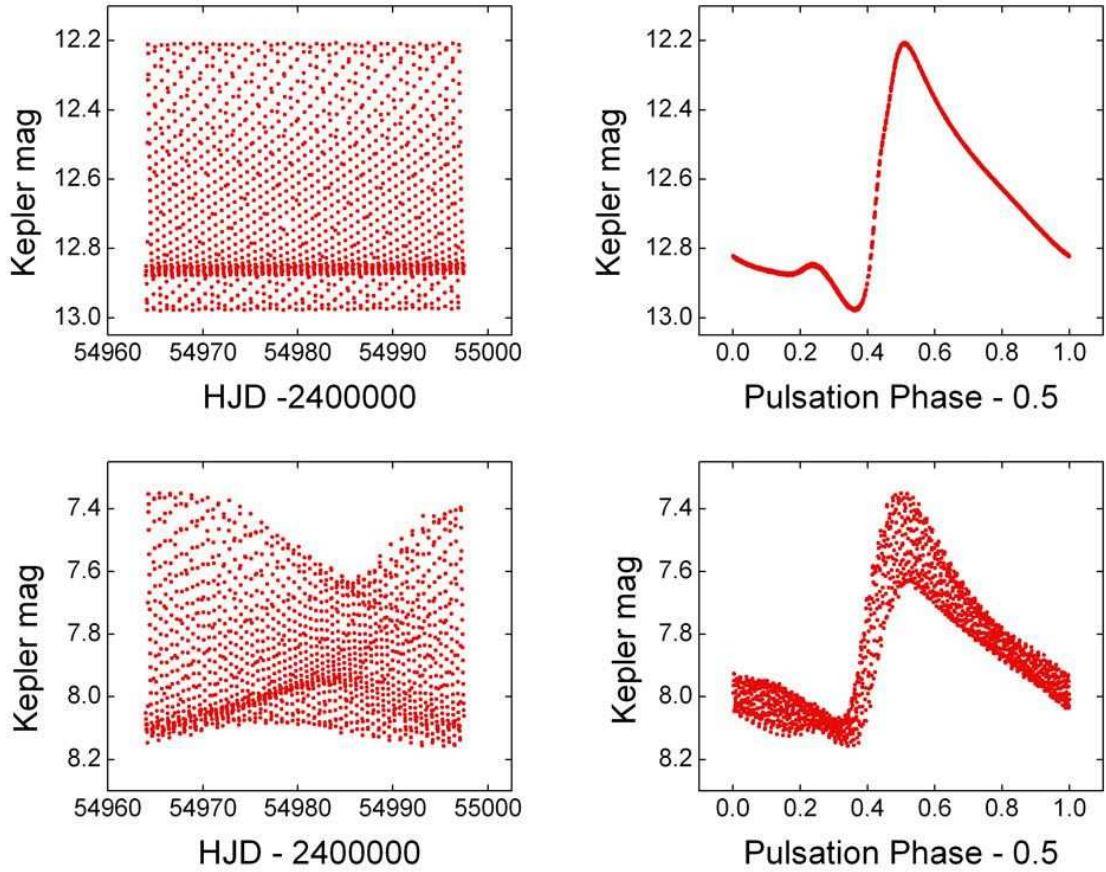}
\caption{
The upper panel shows direct and folded Q1 light curves for the 
non-modulated RR Lyrae star NR Lyrae, while the lower panel illustrates the
rapid evolution of oscillation wave form present in RR Lyrae itself -- characteristic 
of many 
RR Lyrae stars monitored with {\em Kepler}.}
\label{fig:rrlyrae}
\end{figure}

The phenomenon of amplitude and phase modulation of RR Lyr stars -- the
so-called Blazhko effect~\citep{bla07} -- is one of the most stubborn problems of the
theory of radial stellar pulsations.
With {\em Kepler} photometry, we will be able to resolve
important issues, such as period variation, the stability of the pulsation
and modulation, the incidence rate of the Blazhko effect, multiple modulation
periods, and the existence of ultra-low amplitude modulation. Our findings should
constrain existing~\citep{dze04} and future models.
Moreover, {\em Kepler} may find second and higher radial overtones and
possibly nonradial modes \citep{gru07} in the long uninterrupted time series.
Figure~\ref{fig:rrlyrae} shows two striking RR Lyrae light curves from the early {\em Kepler}
data contrasting examples of a constant wave form and strongly modulated one.

\subsubsection{Cepheids}

Classical Cepheids are the most important distance indicators of the nearby
Universe.
Several candidates~\citep{blo09} have been found in the {\em Kepler} field.
Being long-period radial pulsators, Cepheids may show instabilities that have been
hidden in the sparsely sampled, less accurate ground-based observations.
{\em Kepler} also provides an excellent opportunity to follow period change and
link it to internal structure variations due to stellar evolution.
Period variation and eclipses will uncover binary Cepheids;
faint companions may directly affect their role as distance
calibrators~\citep{sza03}.

The discovery of non-canonical Cepheid light variations is highly probable.
Ultra-low amplitude~\citep{sza07} cases are in an evolutionary stage 
entering or leaving the instability strip~\citep{buc02}, 
thus providing information
on the driving mechanism and scanning the previously unexplored domains of
the period-amplitude relation.
Nonradial modes (\citealp{mul07}; \citealp{mos09}), and solar-like oscillations driven by convection, may
provide a valuable tool to sound inner stellar structure, adding yet another
dimension to asteroseismic investigations of Cepheids.

\subsubsection{Slowly Pulsating B and $\beta$ Cep Stars}

Slowly pulsating B (SPB) stars are mid-to-late B-type stars oscillating in
high-order g modes with periods from
0.3 to 3 days. These oscillations are driven by the $\kappa$~mechanism acting in
the iron opacity bump at around 200\,000~K (\citealp{dze93}; \citealp{gau93}).  Since most SPB stars
are multi-periodic, the observed variations have long beat periods and are generally
complex. The large observational efforts required for in-depth asteroseismic studies
are hard to achieve with ground-based observations.

With the ultra-precise {\em Kepler} photometry of SPB stars spanning 3.5 years, it will
be possible to
search for signatures of different types of low-amplitude oscillations to probe
additional internal regions.
{\em Kepler} can resolve the recent suggestion that
hybrid SPB/$\delta$\,Sct stars exist~\citep{deg09}, thus filling the gap
between the classical instability strip and the one for B stars
with nonradial pulsations -- which theory does not predict.
The nature of the excess power recently found in $\beta$ Cep and SPB stars
(\citealp{bel09}; \citealp{deg09}) can be quantified.
{\em Kepler} should also allow us to investigate the stability of the periods
and amplitudes, and to search for evolutionary
effects in the more evolved B stars.
{\em Kepler's} photometric capabilities will help us
detect and explore both the deviations from regular period spacings predicted for
g modes in the asymptotic regime~\citep{mig08} and frequency multiplets induced by stellar
rotation and/or magnetic fields.
Finally, we will
test the developing models that include the effects of rotation on
g mode pulsations.

So far, some $\beta$~Cep stars have been studied asteroseismically
with data from intensive ground-based
observational campaigns~\citep{aer08}.
These studies placed limits on convective core overshooting
and demonstrated the presence of non-rigid internal rotation \citep{aer03},
in addition to identifying constraints on the internal structure,
opacities, and abundances (see, e.g., \citealp{pam04}; \citealp{bri07}; \citealp{das09}).
With the new level of precision provided by {\em Kepler}, we can go further, 
such as examining the suspected presence of solar-like oscillations in 
$\beta$ Cep stars~\citep{bel09}.  The detection of more pulsation modes is key
to detailed analyses of internal rotation. New methods using rotational 
mode splittings and their asymmetries have been developed by \citet{sua09}, and
might help to test theories describing angular momentum redistribution 
and chemical mixing due to rotationally induced turbulence. The analysis 
of ``hybrid" p  and g mode pulsators will result in tighter constraints 
on stellar structure, particularly on opacities~\citep{han09b}.
This increased understanding of $\beta$ Cep stars through 
asteroseismology can provide useful information about the chemical 
evolution of the Universe because $\beta$ Cep stars, pulsating in modes of low radial order and 
periods between about $2 - 7$ hours, are ideal for determining the 
interior structure and composition of stars around 10 $\Msun$---precursors of
type II supernovae.

\subsubsection{Miras and Semiregular Variables}

Miras and semiregular variables (M giants) are the coolest and most luminous
KASC targets, representing advanced evolutionary stages of low- and
intermediate-mass stars such as the Sun. Ground-based photometry has shown many
of these stars to have complicated multi-mode oscillations on timescales
ranging from several hundred days down to ten days and probably shorter (\citealp{woo00}; \citealp{tab09}).
The oscillations are strongly coupled to important, but
still poorly understood mechanisms, such as convection and mass loss.  

The uninterrupted coverage and unprecedented precision of {\em Kepler} photometry will
allow the first application of asteroseismology to M giants.  We anticipate the
first exciting results in this field after the first year of observations, while
the most intriguing theoretical questions will require the full time span of the
project. By measuring the oscillation frequencies, amplitudes and mode
lifetimes we will challenge stellar models and examine the interplay between
convection and the $\kappa$~mechanism, and the roles of both in exciting and
damping the oscillations~\citep{xio07}.  We also hope to shed light on the
mysterious Long Secondary Periods phenomenon~\citep{nic09} and to
search for chaotic behavior and other non-linear effects that arise naturally
in these very luminous objects \citep{buc04}.

\subsection{Oscillating Stars in Binaries and Clusters}

\begin{figure}
\plotone{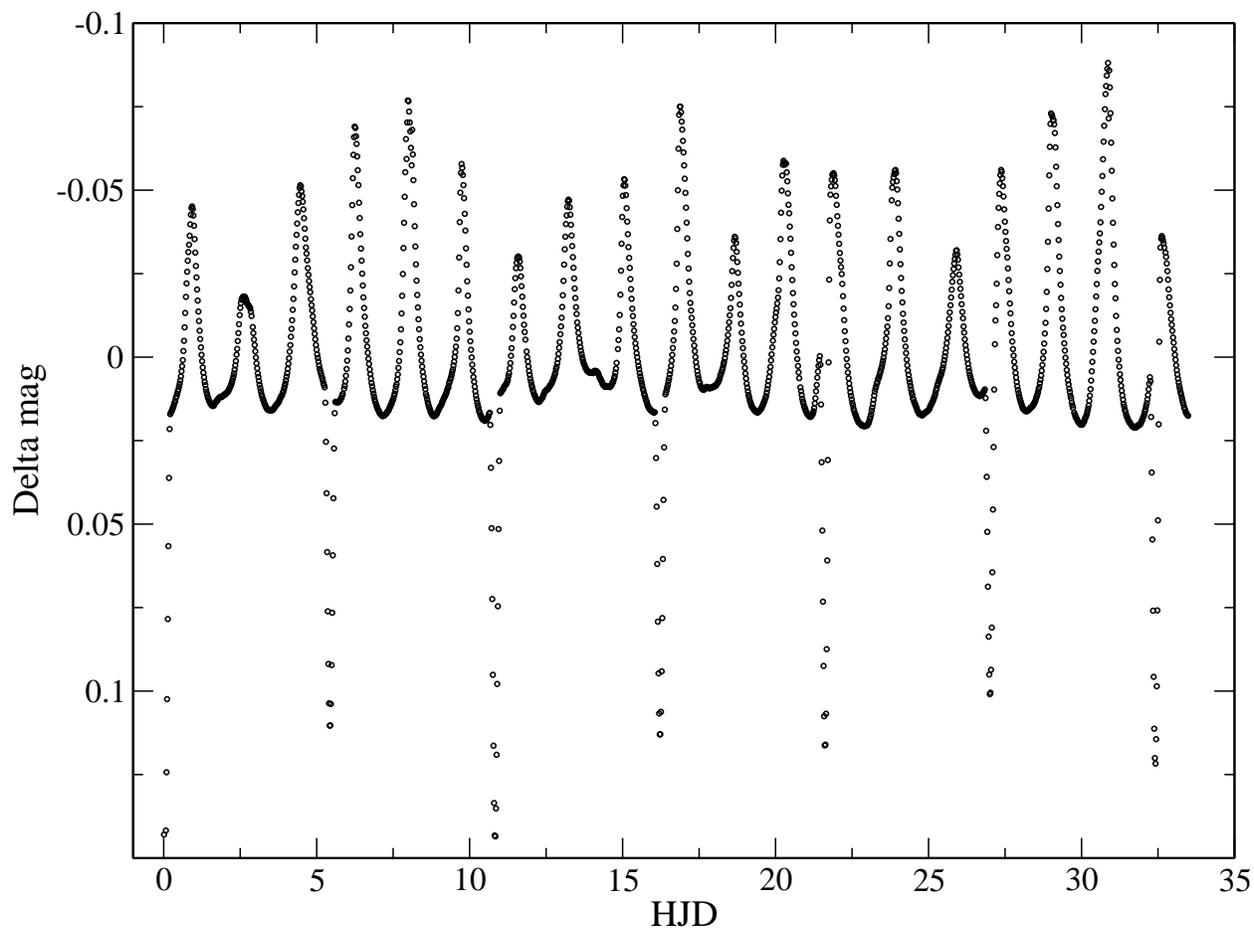}
\caption{
Time series over Q1 of the slowly pulsating B-star, KIC-11285625, which 
shows both large amplitude multi-periodic pulsations, and primary and  
secondary eclipses.  With the addition of ground-based radial-velocity
studies and the exquisite {\em Kepler} light curve elucidating the broad
range of phenomena visible here, this is likely to become one of the 
more important asteroseismic targets.}
\label{fig:SPB}
\end{figure}

One of the goals of the asteroseismology program is to model the
oscillation frequencies of
stars in eclipsing binaries. In order to find such stars in the {\em Kepler}
field of view, a global variability classification treating all KASC stars
observed in Q0 and Q1 was performed with the methodology by \citet{deb09}.  This
revealed
hundreds of periodic variables all over the H-R Diagram, among which
occur stars
with activity and rotational modulation in the low-frequency regime,
Cepheids,
RR\,Lyr stars, and multiperiodic nonradial pulsators~\citep{blo09}.  More than 100 new
binaries were also identified. About half of those are likely to be ellipsoidal variables
while the other half are eclipsing binaries, several of which have pulsating
components. One example for a case unknown prior to launch is shown in
Figure~\ref{fig:SPB},
where we see pulsational behavior which is typical for gravity modes in
a slowly
pulsating B star~\citep{deg09} in an eclipsing system with a period of about 11
days with both primary and secondary eclipses.
Examples of pulsating red giants in eclipsing binaries are clear in 
early {\em Kepler} data~\citep{hek09},
and discussed for the above case of KOI-145.
Eclipsing binaries with pulsating components not only provide stringent
tests of
stellar structure and evolution models, they also offer the opportunity to
discover planets through changes of the pulsation frequencies due to the
light
travel time in the binary orbit~\citep{sil07}.  For these reasons, pulsating eclipsing
binaries are prime targets within KASC.

Star clusters are Rosetta Stones in stellar structure and evolution.
Stars in clusters are believed to have been formed from the same cloud of gas at
roughly the same time, leaving fewer free parameters when analyzed as a
uniform ensemble, which allows stringent tests of stellar evolution
theory. Seismic data enable us to probe the interior of models, and thus
should allow us to test aspects of stellar evolution that cannot be
addressed otherwise, such as whether or not the sizes of convective cores in
models are correct.
Asteroseismology of star clusters has been a long sought
goal that holds promise of rewarding scientific return.  In particular,
the advantages of asteroseismology for clusters are that, unlike estimates of
colors and magnitudes, seismic data do not suffer from uncertainties in
distance or extinction and reddening.

\begin{figure}
\plotone{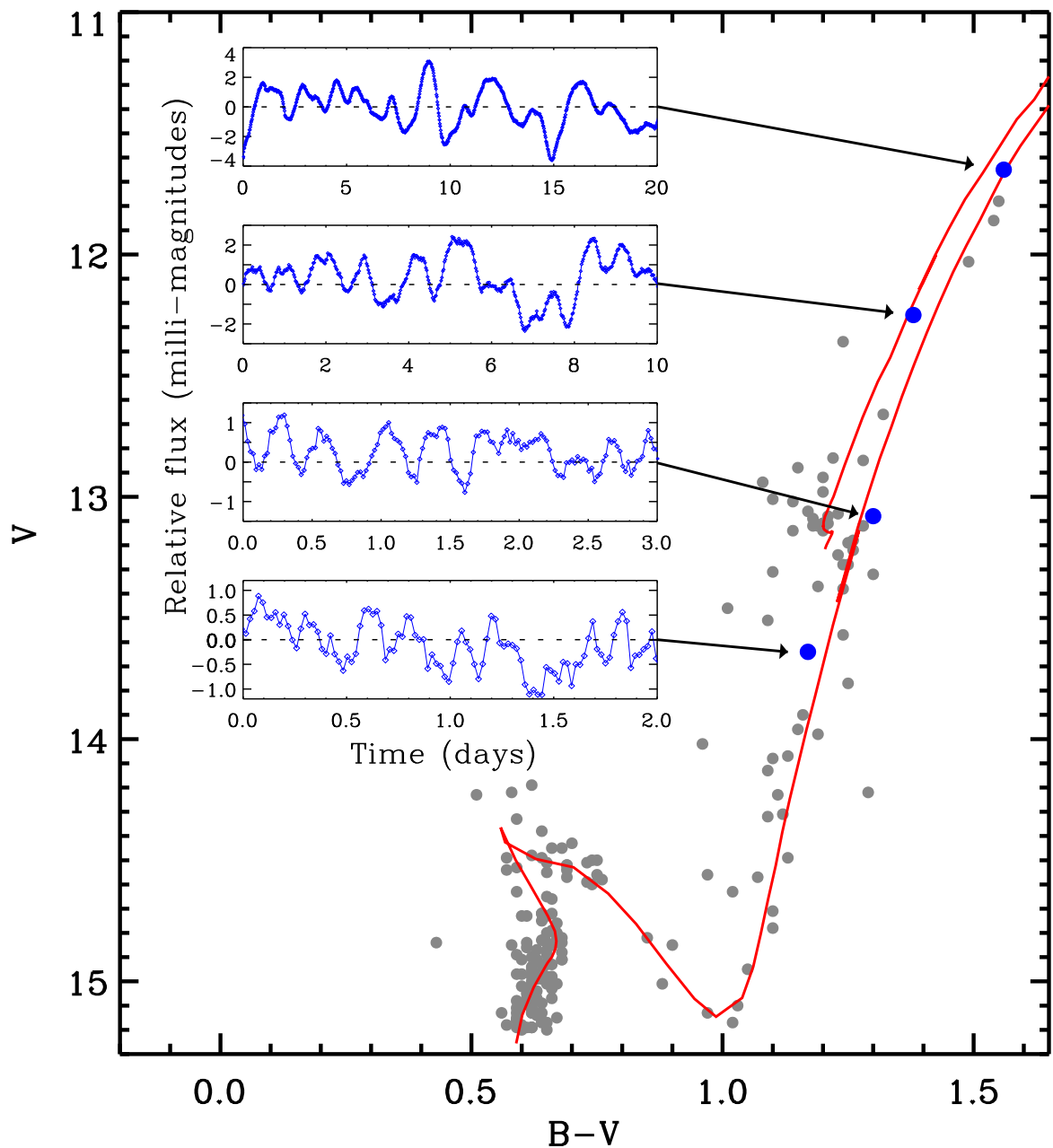}
\caption{
Color-magnitude diagram for NGC~6819. The grey points show stars that have
high radial-velocity probabilities from~\citet{hol09}. The curve (red in on-line version) is a solar metallicity,
2.5 Gyr isochrone of~\citet{mar08}. The larger, dark-grey (blue) points are four of the {\em Kepler} targets; we show
the time series of their brightness fluctuation in the insets. As can be
seen from the inset, the timescale of the oscillations decreases
with increasing apparent magnitude of the stars.}
\label{fig:cluster}
\end{figure}

There are four open clusters in the {\em Kepler} field of view, NGC 6791,
NGC~6811, NGC~6866 and NGC~6819. In Figure~\ref{fig:cluster} we show a color-magnitude
diagram of stars in NGC~6819 and point out the relative flux variation of
four stars along the giant branch. Initial data from {\em Kepler} have
allowed the first clear measurements of solar-like oscillations in cluster stars
by \citet{ste10}. The oscillations of the stars
are clear even without any further analysis. The power
in the oscillations changes with luminosity and effective temperature as
expected from ground-based and early space-based
observations of nearby field stars.  The data further enable us to
determine cluster membership by lining up the power spectra of the observed
stars in order of their apparent magnitude. We have detected several possible
non-members in this manner, which previously all had high membership probabilities
($P>80$\%) from radial-velocity measurements.

While the currently available data from {\em Kepler} are for stars on the giant
branch of the clusters, future observations will also provide data for the
subgiants and main-sequence stars. Having seismic data for stars at various
stages of evolution will allow us to have independent constraints on the
cluster age. Since stars in a cluster are coeval and have the same
metallicity, the process of modelling detailed seismic data of these stars
will allow us to test the various physical processes that govern
stellar evolution.

\section{Summary}

The phenomenal promise of the {\em Kepler Mission} to invigorate stellar
astrophysics through the study of stellar oscillations is clearly
being realized.  The unprecedented combination of temporal coverage and
precision is sure to provide new insights into many classical variable
stars, to increase by more than two orders of magnitude the number of cool
main-sequence and subgiant stars observed for asteroseismology and to
completely revolutionize asteroseismology of solar-like stars.

\acknowledgments

{\em Kepler} is the tenth Discovery mission.  Funding for this
mission is provided by NASA's Science Mission Directorate.
CA, JDR and JD received funding from the European
Research Council under the European Community's Seventh Framework Programme
(FP7/2007--2013)/ERC grant agreement n$^\circ$227224 (PROSPERITY), as well as
from the Research Council of K.U.Leuven (GOA/2008/04) and from
the Belgian Federal Science Policy Office Belspo.
We are grateful to the legions of highly skilled individuals
at many private businesses, universities and research centers
through whose efforts the marvelous data being returned by
{\em Kepler} have been made possible.

{\it Facilities:} \facility{The Kepler Mission}

\newpage

% For your review copy (i.e., the file you initially send in for
% evaluation), you can use the {figure} environment and the
% \includegraphics command to stream your figures into the text, placing
% all figures at the end.  For the final, revised manuscript for
% acceptance and production, however, PostScript or other graphics
% should not be streamed into your compiled file.  Instead, set
% captions as simple paragraphs (with a \noindent tag), setting them
% off from the rest of the text with a \clearpage as shown  below, and
% submit figures as separate files according to the Art Department's
% instructions.

\end{document}